\documentclass{aa}  
\usepackage{txfonts}
\usepackage{hyperref}

\usepackage{graphicx}	
\usepackage{amsmath}	
\usepackage{amssymb}	
\usepackage{xspace}

\newcommand{\chandra}{\textit{Chandra}\xspace}

\newcommand*{\lunit}{\ensuremath{\mathrm{erg\,s^{-1}}}}
\newcommand*{\funit}{\ensuremath{\mathrm{erg\, cm^{-2}s^{-1}}}}

\newcommand{\fSED}{f\mathrm{_{AGN}^{SED}\xspace}}
\newcommand{\fDSFG}{f\mathrm{_{AGN}^{DSFG}\xspace}}
\newcommand{\fLAE}{f\mathrm{_{AGN}^{LAE}\xspace}}
\newcommand{\fHAE}{f\mathrm{_{AGN}^{HAE}\xspace}}

\begin{document} 
\title{
	Fast SMBH growth in the SPT2349--56 protocluster at $z=4.3$
}
\titlerunning{Fast and obscured SMBH accretion in the SPT2349-56 protocluster at $z=4.3$}
\authorrunning{F. Vito et al.}
   \author{F. Vito\thanks{fabio.vito@inaf.it}\inst{1} \and
   	 W. N. Brandt\inst{2,3,4} \and
   	 A. Comastri\inst{1} \and
          R. Gilli\inst{1} \and
R.\,J.~Ivison\inst{5,6,7,8} \and
G. Lanzuisi\inst{1} \and
B.D.~Lehmer\inst{9}\and
I.E.~Lopez\inst{1,10}\and
P. Tozzi\inst{11} \and
C. Vignali\inst{1,10}
          }
 \institute{
INAF -- Osservatorio di Astrofisica e Scienza dello Spazio di Bologna, Via Gobetti 93/3, I-40129 Bologna, Italy
\and
Department of Astronomy \& Astrophysics, 525 Davey Lab, The Pennsylvania State University, University Park, PA 16802, USA
\and
Institute for Gravitation and the Cosmos, The Pennsylvania State University, University Park, PA 16802, USA
\and
Department of Physics, The Pennsylvania State University, University Park, PA 16802, USA
\and
European Southern Observatory (ESO), Karl-Schwarzschild-Strasse~2, D-85748 Garching, Germany
\and
School of Cosmic Physics, Dublin Institute for Advanced Studies, 31 Fitzwilliam Place, Dublin D02 XF86, Ireland
\and
Institute for Astronomy, University of Edinburgh, Royal Observatory, Blackford Hill, Edinburgh EH9 3HJ, UK
\and
ARC Centre of Excellence for All Sky Astrophysics in 3 Dimensions (ASTRO 3D)
\and
Department of Physics, University of Arkansas, 226 Physics Building, 825 West Dickson Street, Fayetteville, AR 72701, USA
\and
Dipartimento di Fisica e Astronomia, Università degli Studi di Bologna, via Gobetti 93/2, 40129 Bologna, Italy
\and
INAF – Osservatorio Astrofisico di Arcetri, Largo Enrico Fermi 5, 50125 Florence, Italy
%
  }

 \date{}
\abstract
{Environment is one of the main physical drivers of galaxy evolution. The densest regions at high-redshift, i.e. $z>2$ protoclusters, are gas-rich regions characterized by high star-formation activity. The same physical properties that enhance star formation in protoclusters are also thought to boost the growth of supermassive black holes (SMBHs), likely in heavily obscured conditions.}
 {We aim to test this scenario by probing the active galactic nucleus (AGN) content of SPT2349--56, a massive, gas-rich, and highly star-forming protocluster core at $z=4.3$ discovered as an overdensity of dusty star-forming galaxies (DSFGs), and comparing the results with the field environment and other protoclusters,}
 {We observed SPT2349--56 with \textit{Chandra} (200 ks), and search for X-ray emission from the known galaxy members. We also perform a spectral energy distribution fitting procedure to derive the physical properties of the discovered AGN. }
  {We detected in the X-ray band two protocluster members, namely C1 and C6, corresponding to an AGN fraction among DSFGs in the structure of $\approx10\%$. This value is consistent with other protoclusters at $z=2-4$, but higher than the AGN incidence among DSFGs in the field environment.  	
  	Both AGN are heavily obscured sources, hosted in star-forming galaxies with $\approx3\times10^{10}\,\mathrm{M_\odot}$ stellar masses. We estimate that the intergalactic medium in the host galaxies contributes to a significant fraction, or even totally, to the nuclear obscuration.  C1, in particular, is a highly luminous ($L_X=2\times10^{45}\,\lunit$) and Compton-thick ($N_H=2\times10^{24}\,\mathrm{cm^{-2}}$) AGN, likely powered by a $M_\mathrm{BH}>6\times10^8\,\mathrm{M_\odot}$ SMBH, assuming Eddington-limited accretion. Its high accretion rate suggests that it is in the phase of efficient growth required to explain the presence of extremely massive SMBHs in the centers of local galaxy clusters. Considering SPT2349--56 and DRC, a similar protocuster at $z=4$, and under different assumptions on their volumes,  we find that gas-rich protocluster cores at $z\approx4$ enhance the triggering of luminous (log$\frac{L_X}{\lunit}=45-46$) AGN by 3--5 orders of magnitude with respect to the predictions from the AGN X-ray luminosity function at similar redshift in the field environment. This result is not merely driven by the overdensity of the galaxy population in the structures.
  }
{Our results indicate that gas-rich protoclusters at high redshift boost the growth of SMBHs, which will likely impact the subsequent evolution of the structures, and thus represent key science targets to  obtain a complete understanding of the relation between environment and galaxy evolution. Dedicated investigations of similar protoclusters are required to definitively confirm this conclusion with higher statistical significance. }


\maketitle
%

\section{Introduction}

According to the hierarchical growth of cosmic structures, dense regions at high redshift collapse and merge into 
the most-massive gravitationally bound objects in the local universe, i.e., galaxy clusters, which are characterized by more evolved galaxy populations than the field environment \citep[e.g.][]{Alberts14}. Therefore,  galaxy evolution must have been accelerated  in their ancestors, i.e., protoclusters \citep[e.g.][]{Overzier16, Chiang17}. In these regions, star formation is efficiently fueled by large amounts of gas infalling from the forming cosmic web \citep[e.g.][]{Umehata19}, and is likely boosted by the high rates of galaxy interactions and mergers in these dense and unvirialized systems \citep[e.g.][]{Liu23}.

Radiative and mechanical feedback produced by gas accretion onto supermassive black holes (SMBHs) observed as active galactic nuclei (AGN) plays a fundamental role in regulating, and eventually hindering, further galaxy and SMBH growth in cluster members \citep[e.g.][]{Fabian12,Gilli19,Gaspari20}. However, the effect of a dense environment on the triggering of nuclear activity at high redshift is still not well understood. X-ray observations are the best tools to investigate the incidence and physical properties of the AGN population in protoclusters, as bright X-ray emission is a reliable and nearly complete tracer of nuclear activity, even in the presence of heavy obscuration \citep[e.g.,][]{Brandt15, Ivison19}. Dedicated X-ray programs with \chandra generally find enhanced AGN activity in protoclusters with respect to the field environment at similar redshift and local galaxy clusters \citep[e.g.,][but see also \citealt{Yang18b, Macuga19}]{Lehmer09b,Lehmer13, Digby-North10, Tozzi22a}. These results support a scenario in which the large reservoirs of gas and the high rate of galaxy interactions promote the growth of SMBHs in the protocluster galaxy members, in addition to boosting the star-formation activity. Theoretical models \citep[e.g.][]{Hopkins06a} predict that these conditions favor fast, efficient, and possibly heavily obscured nuclear accretion. Most protocluster AGN are indeed characterized as being heavily obscured \citep[e.g.][]{Vito20, Monson23}.These properties are typical of the peak phases of SMBH mass building, after which AGN feedback hampers additional galaxy and SMBH growth, eventually impacting the entire cluster’s evolution. In addition, the AGN enhancement may also be an effect of galaxies in protoclusters being typically more massive than in the field environment \citep[e.g.,][]{Monson21}, as luminous AGN are typically found in galaxy with large stellar masses \citep[e.g.][]{Yang17, Yang18}.

Protocluster candidates are identified via detections of overdensities of galaxies, selected in many different ways (see \citealt{Overzier16} and references therein).  The identification of protocluster candidates as overdensities of dusty star-forming galaxies (DSFGs), Ly$\alpha$ emitters (LAEs),or  Lyman- break galaxies (LBGs) are among the most efficient techniques up to  $z\approx8$ \citep[e.g.][]{Laporte22, Morishita23}. Recently, the high angular resolution and sensitivity of ALMA allowed the identification of two extremely massive and star-forming overdensities of DSFGs,the Distant Red Core (DRC) at $z = 4.0$ \citep[e.g.][]{Oteo18,Ivison20}, and SPT 2349--56 at $z = 4.3$ \citep[e.g.][]{Miller18,Hill20,Hill22}, discovered originally by \textit{Herschel} and the South Pole Telescope, respectively \citep{Vieira10, Ivison16} . The cores of these structures extend to a few hundred kpc in projection, and are unique in terms of overdensity, total gas mass, and SFR density. Based on cluster evolutionary models and simulations, \cite{Oteo18}  and \cite{Hill20} argued that DRC and SPT 2349$-$56  are the likely progenitors of $\approx10^{15}\,\mathrm{M_\odot}$ Coma-like clusters in the local universe.  DRC consists of at least 13 spectroscopically identified 
$z \approx4.0$ DSFGs with individual SFRs in the range of $50-3000\,\mathrm{M_\odot\,yr^{-1}}$, for a total $\mathrm{SFR}\approx6500\,\mathrm{M_\odot\,yr^{-1}}$. A total molecular gas mass $M_{H_2}>10^{12}\,\mathrm{M_\odot}$ was estimated from the \mbox{[C I](1--0)} emission lines detected from the protocluster members. 

SPT 2349--56 was identified with ALMA observations as an overdensity of 30 galaxies spectroscopically confirmed at $z\approx4.3$ via the detection of [C II] and CO(4--3) emission lines \citep{Miller18, Hill20, Rotermund21}. 
Among them, 21 objects are detected in sub-mm/mm continuum emission (850 $\mu m$, 1.1 mm, or 3.2 mm; \citealt{Hill20}), and in this paper we refer to them as DSFGs.
The protocluster members are located in a massive core with radius $R\lesssim20$ arcsec ($\lesssim100$ kpc), and two smaller components at $\approx0.9$ arcmin ($\approx0.75$ Mpc) and $\approx3.8$ arcmin ($\approx1.5$ Mpc) from the center of the main overdensity. Numerical simulations predict that the galaxies in the central core of SPT2349--56 will eventually merge into  the brightest cluster galaxy of the descendant structure \citep{Rennehan20}.  Similar arguments as those used for DRC return total values of $SFR\approx8000\,\mathrm{M_\odot\,yr^{-1}}$ and $M_{vir}\approx10^{13}\,\mathrm{M_\odot}$. Notably, the derived SFR density is  a factor of ten larger than the most extreme values found in simulations at the same redshift (Hill et al. 2020). A total gas mass of $M_{gas}\gtrsim3\times10^{11}\,\mathrm{M_\odot}$ is estimated from the luminosity of the carbon monoxide emission lines \citep{Hill20}.

Using deep optical/near-IR observations, \cite{Apostolovski23} and \cite{Rotermund21}
 identified  additional  members of the SPT2349--56 core as LAEs (9 spectroscopically confirmed galaxies, although 6 of them are marked as ``tentative" in that work) 
 and LBGs (4 galaxies),\footnote{We note that \cite{Rotermund21} selected tens of LBG candidates over a large area (i.e., up to a radius of $\approx2$ arcmin) centered on SPT2349--56. Following \cite{Rotermund21}, we considered only the four galaxies found in the core region as potential members of the protocluster.} respectively. One of the tentative LAE is the counterpart of a sub-mm continuum detected galaxy with neither [C II] nor CO(4--3) emission in \cite{Hill20}, increasing the number of spectroscopically identified DSFGs in the structure to 22. Among the four LBGs in the protocluster core, two are likely counterparts of the DSFGs, named C2 and C17 \citep{Hill22}, one has been found by \cite{Rotermund21} to be a weak [C II] emitter not included in the \cite{Hill20} sample, and one is the counterpart of a LAE.
 Thus, accounting for the few galaxies selected with multiple methods, the total number of individual and spectroscopically confirmed protocluster members is 38.

\cite{Chapman23} detected bright radio emission ($L_\mathrm{1.4GHz,rest}=(2.2\pm0.2)\times10^{26}\,\mathrm{W\,Hz^{-1}}$) with spectral index $\alpha=-1.45\pm0.16$, where $F_\nu\propto\nu^\alpha$, from the inner region of the SPT2349--56 core, likely of AGN origin. The spatial resolution of the radio observations prevented the secure identification of the optical/IR couterpart, as a few protocluster members are consistent with being the host of the radio source, and none of them shows clear AGN features at other wavelengths. Still, mainly based on its large mass, \cite{Chapman23} proposed that one such galaxies, referred to as C6 in our work following the \cite{Hill20} naming convention, is the AGN host.

Due to their extreme properties, these two protoclusters, at similar redshift and identified via similar selection techniques, are unique testbeds to study the link between the availability of huge reservoirs of gas in high-redshift overdense environments, and SMBH growth in the galaxy members. \cite{Vito20} used \chandra observations (140 ks) to investigate the AGN content of DRC, and identified two obscured AGN among 13 DSFGs. These are the two brightest, gas-rich, most strongly star-forming members of the protocluster, and are possibly in a merger phase, as derived from a high angular-resolution ALMA observation of one of them. In particular, the X-ray brightest AGN  ($L_{2-10\,\mathrm{keV}}=2.7_{-1.8}^{+8.9}\times10^{45}\,\mathrm{erg\,s^{-1}}$), namely DRC-2, has remarkable properties. It is as X-ray luminous as optically selected QSOs at all redshifts, but, in contrast to those, obscured by Compton-thick gas column densities, similar to, but even more extreme than, other populations of luminous obscured QSOs such as hot dust-obscured galaxies \citep[e.g.,][]{Vito18b}. 

In this work, we present new \chandra observations (200 ks) of SPT2349--56. Our goals are to 
probe the population of AGN and their physical properties in the extremely gas-rich and dense environment of SPT 2349–-56, and to 
 study more generally the effect of an overdense environment on SMBH growth in the early universe by comparing the AGN content in $z\approx4$ gas-rich protoclusters with lower redshift structures and blank fields.
 Errors are reported at 68\% confidence levels, while limits are given at 90\% confidence levels. We refer to the $0.5-2$ keV, $2-7$ keV, and $0.5-7$ keV energy ranges as the soft band, hard band, and full band, respectively. We assume solar metallicities and
 abundances \citep{Anders89}, and adopt a flat cosmology with $H_0=67.7\,\mathrm{km\,s^{-1}}$ and $\Omega_m=0.307$ \citep{Planck16}.

\section{Data analysis}\label{data_analysis}

In this section, we describe the reduction of the \chandra observations (\S~\ref{data_reduction}) and the source detection procedure (\S~\ref{source_detection}).

\subsection{Data reduction}\label{data_reduction}
We observed the SPT2349--56 protocluster with \chandra for a total of 200 ks, split among 9 pointings (see Tab.~\ref{Tab_obs}). The protocluster core was placed at the aimpoint of the ACIS-S detector, where the \chandra sensitivity is maximum,  and all of the confirmed or candidate protocluster members are covered by the back-illuminated S3 chip. We reprocessed the \chandra\, observations with the \textit{chandra\_repro} script in CIAO 4.15 \citep{Fruscione06},\footnote{\url{http://cxc.harvard.edu/ciao/}} using CALDB v4.10.4,\footnote{\url{http://cxc.harvard.edu/caldb/}} and setting \textit{check\_vf\_pha=yes}, since observations were taken in Very Faint mode.

In order to correct the astrometry of the \chandra observations, we performed source detection on each pointing with the \textit{wavdetect} script with a significance threshold of $10^{-6}$, and then used the  \textit{wcs\_match} and \textit{wcs\_update} tools to compute and apply the astrometric offsets with respect to a reference catalog.
First, we applied a relative astrometric correction to each pointing using OBSID 25267 (i.e., the pointing with the longest exposure,  see Tab.~\ref{Tab_obs}) as reference. Only point-sources with PSF size $\leq3$ arcsec and with $\geq5$ detected counts are considered. Then, we mapped the individual observations onto OBSID 25267 and merged all of the observations with the \textit{reproject\_obs} tool. We repeated the source-detection procedure on the merged dataset, but this time we matched the detected point sources to the GAIA DR3 catalog \citep{GAIADR3}\footnote{\url{https://www.cosmos.esa.int/web/gaia/dr3}} , in order to derive and apply the absolute astrometric correction factors. Only two X-ray point sources could be matched with GAIA objects on the ACIS-S3 chip, and the average spatial offset is 0.15 arcsec. This value can be considered the systematic spatial uncertainty of the X-ray dataset.

We obtained images and exposure maps with the \textit{reproject\_obs} tool, while we extracted spectra, response matrices, and ancillary files of the detected sources (see \S~\ref{source_detection}) from individual pointings using the \textit{specextract} tool and added them using the \textit{mathpha}, \textit{addrmf}, and \textit{addarf} HEASOFT tools\footnote{\url{https://heasarc.gsfc.nasa.gov/docs/software/heasoft/}}, respectively, weighting by the individual exposure times. Ancillary files, which are used to derive fluxes and luminosities, were aperture corrected by setting via the \textit{correctpsf} parameter in \textit{specextract}.

\begin{table}
	\caption{Summary of the \chandra\, observations of SPT2349--56.}
	\begin{tabular}{ccccccccc} 
		\hline
		\multicolumn{1}{c}{{ OBSID}} &
		\multicolumn{1}{c}{{ Start date }} &
		\multicolumn{1}{c}{{ $T_{exp}$ [ks]}} \\ 
		25267 &2023-06-13 &36\\
		25704 &2023-09-21 &30\\
		25705 &2023-08-25 &30\\
		25706 &2023-02-13 & 14\\
		25707 &2023-08-07 & 32\\
		25708 &	2023-04-20 & 14\\
		27712&2023-04-14 & 16\\
		27804&2023-04-23 & 14\\
		27904&2023-06-18 & 13\\
		\hline
	\end{tabular} \label{Tab_obs}\\
\end{table}

\begin{figure*}
	\begin{center}
		\hbox{
			\includegraphics[width=180mm,keepaspectratio]{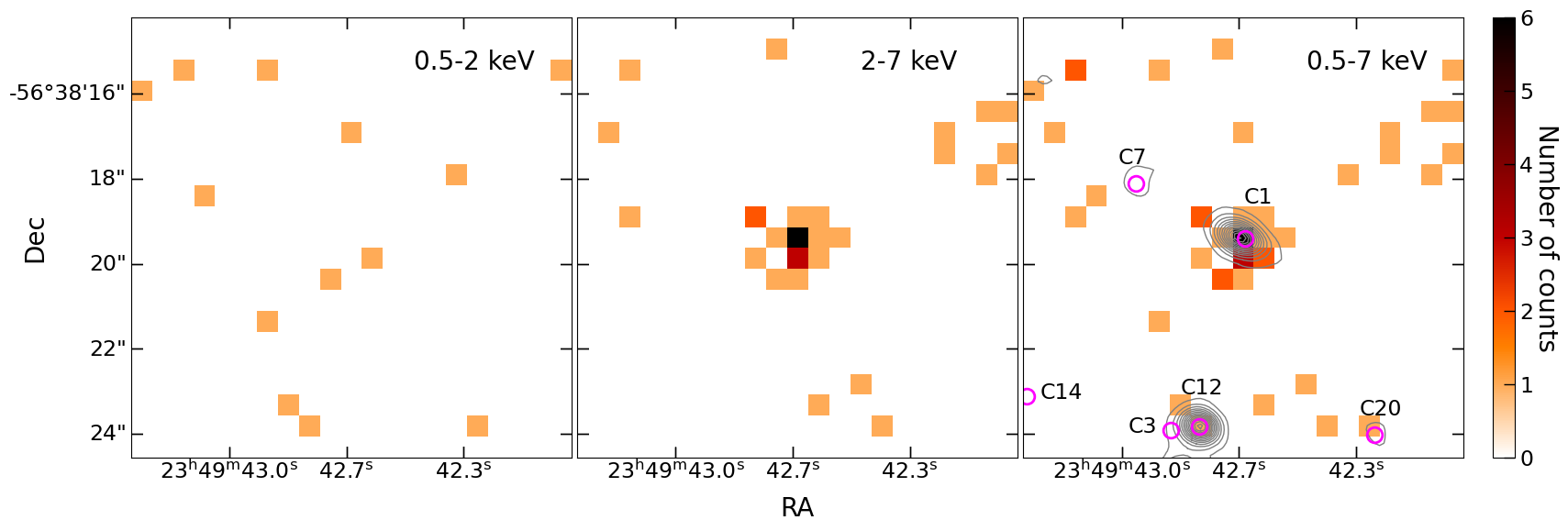} 
		}
	\end{center}
	\caption{From left to right, soft-band, hard-band, and full-band \chandra images ($10^{\prime\prime}\times10^{\prime\prime}$) of C1.  In the right panel, we also plot the ALMA Band 7 continuum contours (in grey) and mark the positions (purple circles) of spectroscopically identified protocluster members. The contours have been derived from the reduced ALMA data of \cite{Hill20} with beam size of $0.35^{\prime\prime}\times0.29^{\prime\prime}$, and start at $5\sigma$ with steps of $7\sigma$ . 
	}\label{Fig_C1_Xray}
\end{figure*}

\begin{figure*}
	\begin{center}
		\hbox{
			\includegraphics[width=180mm,keepaspectratio]{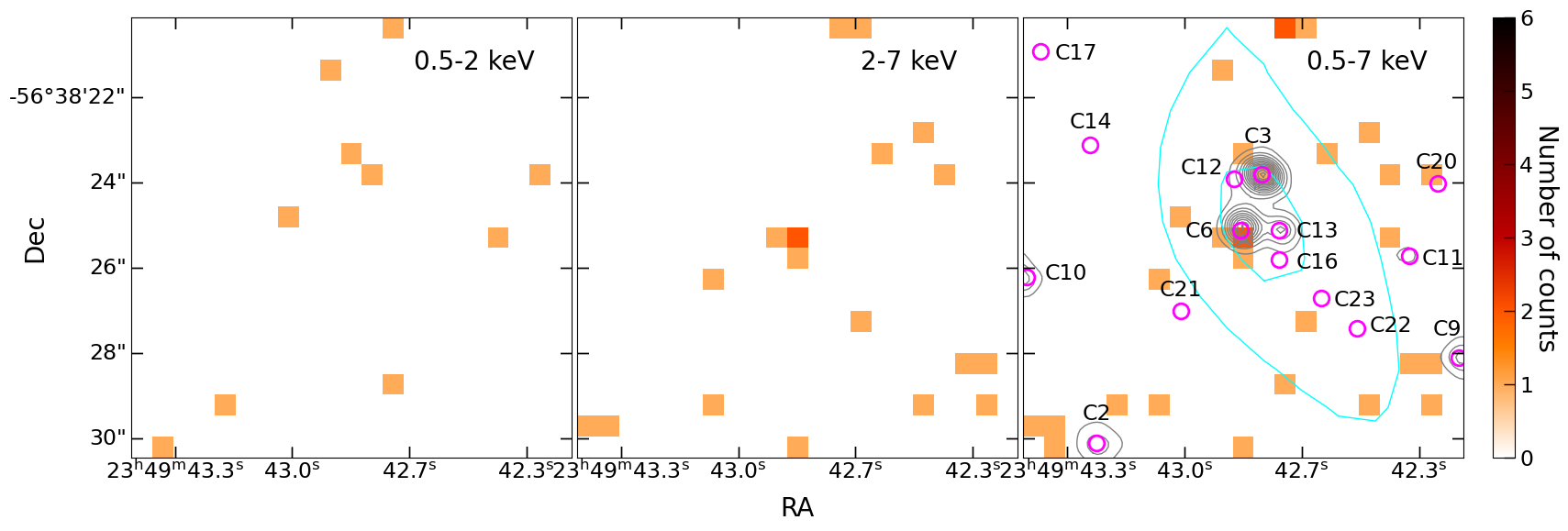} 
		}
	\end{center}
	\caption{Same as Fig~\ref{Fig_C1_Xray}, but for source C6. We added in cyan the ATCA 2.2 GHz contours (beam size of $7.7^{\prime\prime}\times4.2^{\prime\prime}$), starting at $3\sigma$ with steps of $3\sigma$, obtained from the data reduced and presented by \cite{Chapman23}.
	}\label{Fig_C6_Xray}
\end{figure*}

\subsection{Source detection and X-ray photometry}\label{source_detection}

We assessed the detection of the SPT2349--56 protoclusters member candidates \citep{Hill20, Hill22, Rotermund21}  in the soft, hard, and full bands using the binomial no-source probability \mbox{\citep{Weisskopf07,Broos07}}
\begin{equation}
	P_B(X\geq S)= \sum_{X=S}^{N}\frac{N!}{X!(N-X)!}p^X(1-p)^{N-X},
\end{equation}
where $S$ is the total number of counts in the source region in the considered energy band, $B$ is the total number of counts in the background region, \mbox{$N=S+B$}, and $p=1/(1+BACKSCAL)$, with $BACKSCAL$ being the ratio of the background and source region areas. The source counts are extracted from  circular regions with $R=1''$, whereas the background counts are measured from nearby regions free of evident X-ray sources. We checked that reasonably different choices of extraction regions returned consistent results. We defined $(1-P_B)>0.99$ as the detection threshold, a value often used to assess the X-ray detections of objects with pre-determined positions \citep[][]{Vito19b,Vito20}. Due to the small projected distances between the protocluster members, especially in the core of the structure, the extraction regions may overlap. In such cases, we assign each detected photon to the nearest galaxy, to avoid double counting.

\begin{table*}
	\caption{Positions and X-ray photometric properties of the two members of the protocluster detected with \chandra (see \S~\ref{source_detection}). }
	\begin{tabular}{ccccccccccc} 
		\hline
		\multicolumn{1}{c}{{ ID }} &
		\multicolumn{1}{c}{{ RA}} &
		\multicolumn{1}{c}{{ DEC }} &
		\multicolumn{1}{c}{{ C$_{\mathrm{sb}} $}} &
		\multicolumn{1}{c}{{ C$_{\mathrm{hb}} $}} &
		\multicolumn{1}{c}{{ C$_{\mathrm{fb}}$ }}  &
		\multicolumn{1}{c}{{ HR }}  &
		\multicolumn{1}{c}{{ $\Gamma_{eff}$ }}  &
				\multicolumn{1}{c}{{ $N_H $ }} &
				\multicolumn{1}{c}{{ $F_{0.5-7\,\mathrm{keV}}$ }}&
	\multicolumn{1}{c}{{ $L_{2-10\,\mathrm{keV}}$ }}
		\\
				& J2000& J2000 &  & & & &  &$10^{24}\,\mathrm{cm^{-2}}$ & $10^{-15}\,\funit $& $10^{45}\,\lunit$\\
		(1) & (2) & (3) & (4) & (5) & (6) & (7) & (8) & (9) & (10) & (11) \\
		\hline
		C1 & 23:49:42.65 & -56:38:19.4 &  $<4.8$ &  $19.1^{+ 4.8}_{-4.1}$ & $20.5^{+ 5.0}_{-4.4}$  & $>0.58$ & $<0.13$ & $2.4_{-1.2}^{+2.3}$ & $2.4_{-1.5}^{+4.9}$ & $2.2_{-1.4}^{+4.5}$\\
		C6 & 23:49:42.84 & -56:38:25.1 & $<2.3$ &  $3.6^{+ 2.4}_{-1.7}$ & $3.3^{+ 2.4}_{-1.7}$ & $>0.18$ & $<1.22$ & $>0.4$ & $0.3_{-0.1}^{+0.2}$ & $>0.02$\\
		\hline

		\hline
	\end{tabular} \label{Tab_info}\\
	\tablefoot{(1) ID, (2) RA, and (3) Dec from \cite{Hill20}  of the detected X-ray sources; (4) soft-band, (5) hard-band, and (6) full-band net counts; (7) hardness ratio and (8) corresponding effective photon index; (9) column density; (10) full-band flux; (11) intrinsic luminosity. The column density, flux, and luminosity of C1 are derived via a spectral analysis (Sec.~\ref{C1_spec_analysis}), while these quantities are based on the hardness ratio for C6 (Sec.~\ref{C6_spec_analysis}). }
\end{table*}

Two protocluster members are detected significantly in X-rays: sources C1 ($1-P_B>0.999$ in both the hard and full bands) and C6 ($1-P_B=0.999$ and 0.994 in the hard and full band, respectively), which are among the spectroscopically identified DSFGs in the core region of SPT2349--56 \citep{Hill20}.  Fig.~\ref{Fig_C1_Xray} and \ref{Fig_C6_Xray} present X-ray cutouts of these two sources.

Net counts, which are reported in Tab.~\ref{Tab_info},  were computed in the same $R=1''$ region used for the detection for C6, whereas we used a larger region with  $R=1.5''$ for C1, given its relatively bright emission. We also derived the hardness ratios $HR=\frac{(H-S)}{H+S}$, where $S$ and $H$ are the net counts in the soft and hard band, respectively, and the corresponding effective power-law photon indices, following the procedure of \cite{Vito19b}, which are also  reported in Tab.~\ref{Tab_info}.

We visually inspected all of the remaining secure protocluster members or member candidates presented by \cite{Hill20, Hill22, Rotermund21, Apostolovski23}. Three X-ray photons clustered on three contiguous pixels on top of the galaxy named LBG2 in \cite{Rotermund21} are detected in the full band, suggesting that this object might be a sub-threshold X-ray source.

\section{Results}\label{results}
In this section, we report the results obtained from the X-ray observations of SPT2349--56. In \S~\ref{C1_spec_analysis} and \S~\ref{C6_spec_analysis} we present the results of a spectral analysis of the two detected X-ray sources, C1 and C6, respectively. In \S~\ref{SED_fitting}, we used the X-ray emission of these two galaxies together with the available optical-to-mm photometry to estimate the physical parameters of the AGN host galaxies via a spectral energy distribution (SED) fitting procedure.
In \S~\ref{stacking} we investigate possible evidence of low-rate SMBh accretion in the individually undetected galaxies of the structure via an X-ray stacking analysis.

\subsection{X-ray spectral analysis of C1}\label{C1_spec_analysis}

\begin{figure}
		\includegraphics[width=90mm,keepaspectratio]{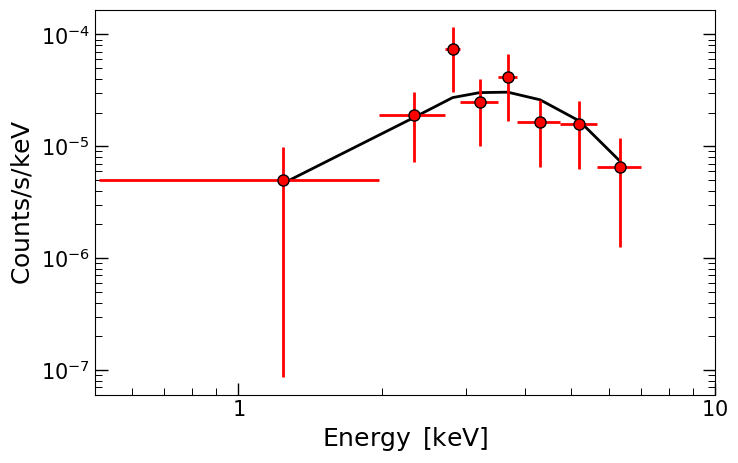} 
		\includegraphics[width=90mm,keepaspectratio]{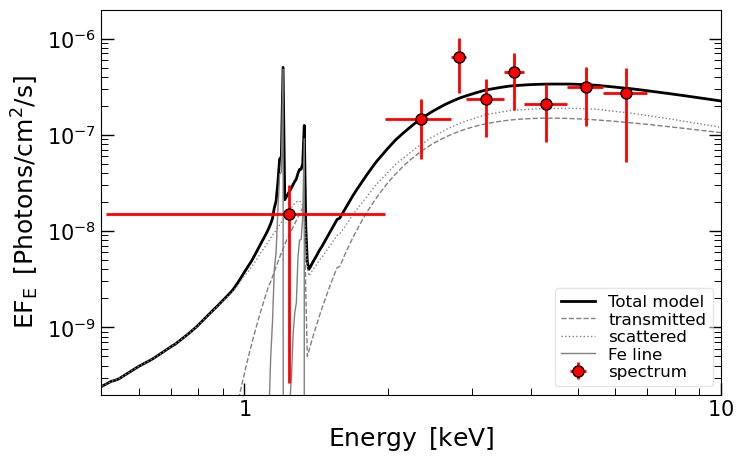} 
	\caption{Top panel: observed spectrum (red circles) and best-fitting MYTorus model (black line) of C1 (see \S~\ref{C1_spec_analysis}). Bottom panel: response-corrected spectrum and best-fitting model. We also show the individual additive components of the model with grey lines, as reported in the legend. In both panels, the spectrum is binned at $1\sigma$ for display purposes. }\label{Fig_C1_spectrum}
\end{figure}

The relatively bright hard-band detection of C1 with no associated soft-band emission and the resulting flat effective photon index (Tab.~\ref{Tab_info}) are strongly indicative of the presence of a large column density of gas ($N_H$) obscuring this high-redshift galaxy. We performed a basic spectral analysis with XSPEC v.12.13 \citep[][]{Arnaud96}\footnote{We used the $W$-statistic, which is suitable in case of background-subtracted spectra with low number of counts. See \url{https://heasarc.gsfc.nasa.gov/xanadu/xspec/manual/XSappendixStatistics.html} and \cite{Cash79}.} to measure physical quantities, such as the $N_H$, the observed fluxes, and the intrinsic luminosity. Following the analysis of DRC-2 \citep{Vito20}, we used the MYTorus model \citep{Murphy09}, accounting for the Galactic absorption \citep{Kalberla05}, and fixing $\Gamma=1.9$, the normalizations of the scattered and line components to that of the transmitted component, and the inclination angle $\Theta=90$ degrees. Therefore, the only two parameters left free to vary were $N_H$ and the intrinsic powerlaw normalization.  Fig.~\ref{Fig_C1_spectrum} reports the observed and response-corrected spectrum and best-fitting model, highlighting the hardness of the source, as typically found for heavily obscured AGN.

\begin{figure}
			\includegraphics[width=90mm,keepaspectratio]{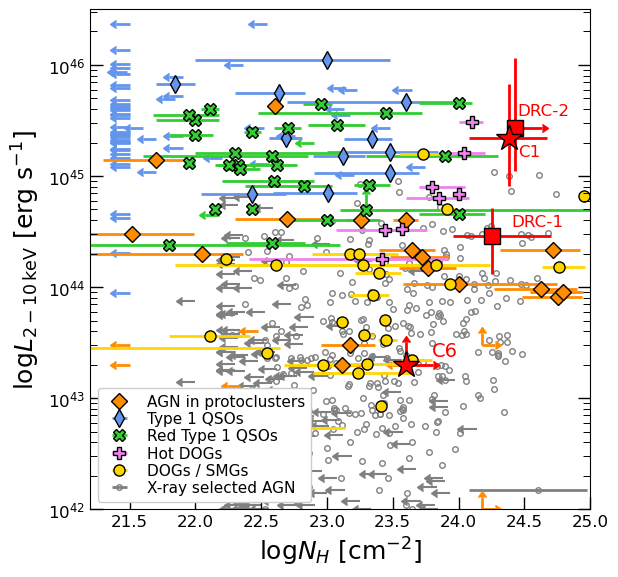} 
	\caption{X-ray luminosity versus column density for different populations of AGN: optically selected Type-1 QSOs at $z=1.4-4.5$ (blue symbols; from \citealt{Just07,Martocchia17}), dust-reddened QSOs at $z=0.4-3.2$ (green symbols; from \citealt{Urrutia05, Banerji14, Mountrichas17, Goulding18, Lansbury20}), DSFGs at $z=0.3-5.2$ (yellow symbols; from \citealt{Wang13_1, Corral16, Zou20}), Hot DOGs at at $z=1.0-4.6$  (violet symbols; from \citealt{Stern14, Assef16, Ricci17, Vito18b, Zappacosta18}), X-ray selected AGN in the Spiderweb and SSA protoclusters \citep[$z=2.2-3.2$, orange symbols,][]{Ivison19, Tozzi22a,Monson23},and X-ray selected AGN in the the \textit{Chandra} deep field-south at all redshifts (gray symbols, with median error bar showed in the bottom-right corner of the plot; from  \citealt{Li19}). The X-ray selected AGN in SPT2349--56 and DRC are plotted as red stars and squares, respectively.
		C1  and DRC-2 \citep{Vito20} have remarkably similar physical properties. Their X-ray luminosities are similar to those of luminous optically selected QSOs, but are obscured by gas column densities even thicker than Hot DOGs.   }\label{Fig_C1_Lx_NH}
\end{figure}

Our best-fitting model returns $N_H=2.4_{-1.2}^{+2.3}\times10^{24}\,\mathrm{cm^{-2}}$, implying that C1 is a Compton-thick AGN. The observed flux $F_{0.5-7\,\mathrm{keV}}=2.4_{-1.5}^{+4.9}\,\times10^{-15}\funit$ corresponds to an absorption-corrected, rest-frame luminosity $L_{2-10\,\mathrm{keV}}=2.2_{-1.4}^{+4.5}\times{10^{45}}\,\lunit$, that is, $\approx1$ dex larger than the break luminosity of the X-ray luminosity function at that redshift \citep[e.g.,][]{Ueda14, Aird15, Vito18a}.  
In Fig.~\ref{Fig_C1_Lx_NH}, we compare the column density and luminosity of C1 from the best-fitting model with other populations of AGN over wide redshift ranges. C1 has an X-ray luminosity typical of bright optically selected blue and red QSOs, which, however, are typically unobscured or, at most, obscured by Compton-thin column densities of gas \citep[e.g.][]{Just07,Martocchia17,Lansbury20}. C1 is even more obscured than hot dust-obscured galaxies (Hot DOGs; e.g., \citealt{Stern14, Vito18b}), which are often considered as representative of an extreme phase of galaxy and SMBH growth, while X-ray selected Compton-thick  AGN and DSFGs have significantly lower luminosities. Intriguingly, C1 and DRC-2, both selected as luminous and obscured AGN in $z\approx4$ gas-rich protoclusters, share the same position in Fig.~\ref{Fig_C1_Lx_NH}.  Only two other AGN with similar X-ray luminosities have been discovered in other protoclusters \citep{Ivison19, Tozzi22a}, but they are unobscured or, at most, mildly obscured, sources. In particular, \cite{Ivison19} detected broad H$\alpha$ emission from 
 the luminous AGN hosted in HATLAS J084933.4+021443, a DSFG in a $z=2.41$ protocluster, which is surprising considering the expected high extinction in such a dust-rich galaxy.

The X-ray luminosity of C1 translates into a bolometric luminosity  $L_\mathrm{bol}\approx8\times10^{46}\,\lunit$ according to the bolometric correction of \cite{Duras20}.
Assuming that the SMBH powering C1 is accreting at the Eddington limit, we can place a lower limit on its mass of $M_\mathrm{SMBH}\approx6.5\times10^8\,\mathrm{M_\odot}$ (but see the caveats raised by \citealt{King24} about this approach). Therefore, either C1 is accreting at super-Eddington rate, or it has already accumulated a significant fraction of the SMBH mass characterizing AGN in the centers of local galaxy clusters. Similar conclusions were also drawn for DRC-2 by \cite{Vito20}. 

We note that the spectrum of C1 can be fitted equally well with a pure reflection model by multiplying the intrinsic powerlaw continuum by a constant equal to zero. In this case, we obtain a similar value of $N_H$ and a factor of $\approx2$ higher luminosity. We also tried leaving $\Gamma$ free to vary, but the fit cannot constrain its value.

\subsubsection{A note on the possible foreground nature of the X-ray emission}
\cite{Rotermund21} reported the spectroscopic identification of a foreground galaxy at $z=2.54$ along the line of sight of C1. Based on considerations on the blue optical colors, magnitudes, and [O III] 5007$\mathrm{\AA}$ emission-line width, they provide an upper limit on its stellar mass of $1.6\times10^9\,\mathrm{M_\odot}$. 
If this were  the host galaxy, the spectral analysis would return
$N_H=1.1_{-0.5}^{+0.6}\times10^{24}\,\mathrm{cm^{-2}}$ and $L_{2-10\,\mathrm{keV}}=4.1_{-2.2}^{+5.4}\times{10^{44}}\,\lunit$. Given the well-known relation between AGN actvity and stellar mass of the host galaxies \citep[e.g.]{Xue10, Lusso11, WangT17, Yang17, Yang18}, it is highly unlikely that such a small object host a moderately luminous AGN. Moreover, its rest-frame UV spectrum present no indications of AGN activity \citep{Rotermund21}. Therefore, in this paper we assume that the X-ray source is hosted by the DSFG at $z=4.3$.

\subsection{X-ray spectral properties of C6}\label{C6_spec_analysis}

Due to the low number of detected X-ray counts, we did not attempt to perform a spectral fit of C6. Instead, we assumed a simple $\Gamma=1.9$ power-law as intrinsic spectrum, and estimated $N_H>4.0\times10^{23}\,\mathrm{cm^{-2}}$ in order to reproduce the observed value of hardness ratio, accounting for the proper instrumental response and PSF correction by using the ancillary and response file extracted at the position of C6 (see Sec.~\ref{data_reduction}). We used this model 
to estimate the observed flux and the intrinsic luminosity of C6 (Tab.~\ref{Tab_info}). 
Based on these loose constraints, we conclude  that C6 confidently is a heavily obscured AGN, although possibly not as extreme as C1.

The X-ray detection of C6 strongly supports that this galaxy is the host of the radio AGN discovered by \cite{Chapman23} with . The ATCA 2.2 GHz contours presented by \cite{Chapman23} are reported in Fig.~\ref{Fig_C6_Xray} in cyan. The limit on the X-ray luminosity of C6 is consistent with the relation between radio and X-ray emission from radio-loud AGN \citep[e.g.,][]{Fan16, DAmato20, Mazzolari24}.

\begin{figure*}
		\includegraphics[width=90mm,keepaspectratio]{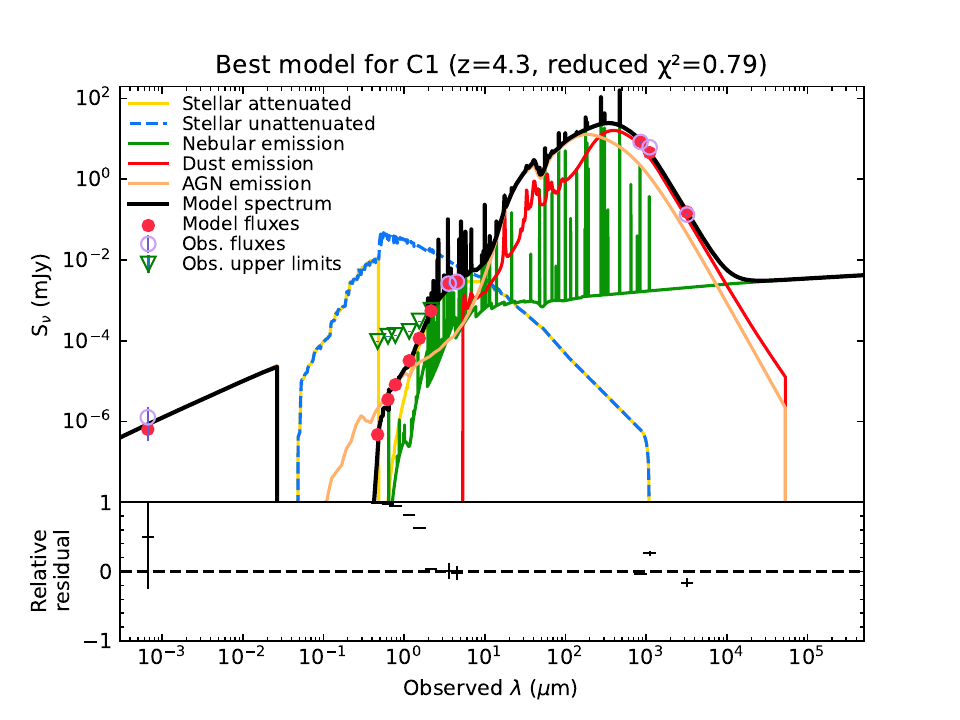} 
		\includegraphics[width=90mm,keepaspectratio]{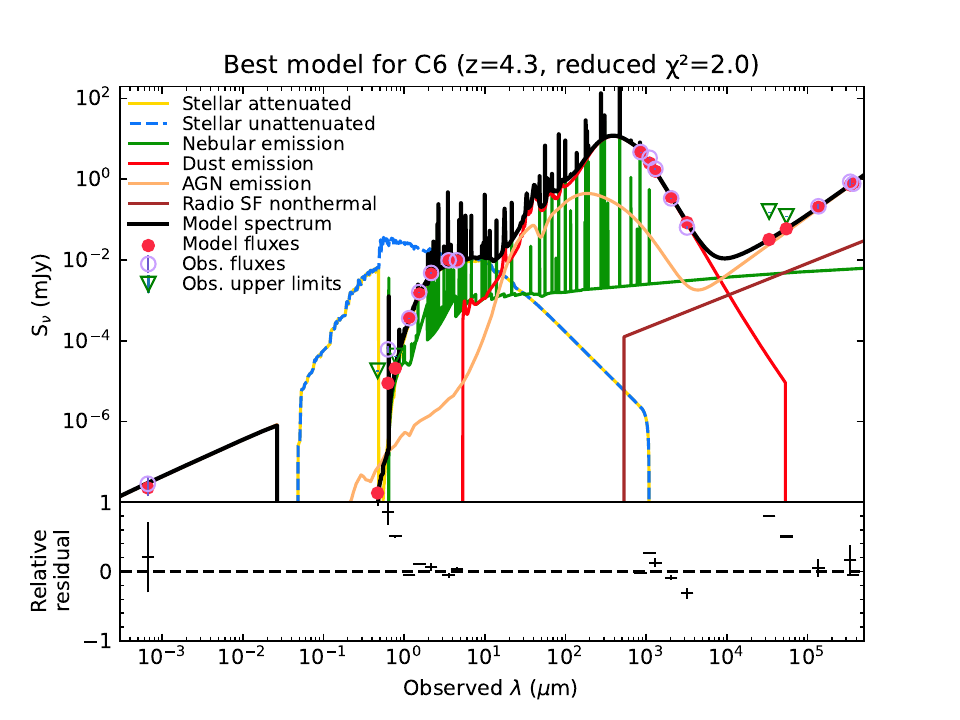}
	\caption{Best-fitting models for C1 and C6 returned by the CIGALE fitting code. The optical-to-mm photometric points have been collected from \cite{Hill20,Hill22}, while for C6 we added the radio measurements of \cite{Chapman23}. The absorption-corrected X-ray fluxes correspond to the X-ray models discussed in Sec.~\ref{C1_spec_analysis} and \ref{C6_spec_analysis}. The sub-mm/mm photometry of C1 is dominated by the AGN reprocessed emission. The AGN in C6 contributes more modestly at such wavelengths, but dominates the radio emission.}\label{Fig_SED}
\end{figure*}

\subsection{SED fitting}\label{SED_fitting}

We used CIGALE v2022.1 \citep[e.g.][]{Boquien19,Yang22}  to fit the  SEDs  of the two X-ray selected AGN in SPT2349--56. Fig.~\ref{Fig_SED} presents the best-fitting SED models, and the resulting physical parameters are reported in Tab.~\ref{Tab_SED_results}. CIGALE produces a SED model for every combination of the input parameters, convolves it with the filters corresponding to the utilized photometric points, and computes the likelihood exp$(-\chi^2/2)$ of every model in a Bayesian framework. Then, it computes the marginalized probability distribution function of each physical parameter based on the likelihood of all models, and returns the mean and the standard deviation, which can be considered as the estimated value and associated uncertainty.

We used the optical-to-mm photometric points from Gemini-GMOS, FLAMINGOS-2, \textit{HST}-WFC3, \textit{Spitzer}-IRAC, and \textit{ALMA} presented by \cite{Hill20,Hill22,Chapman23}. Upper limits at $3\sigma$ are adopted here for filters in which the sources are not detected. 
We fitted the SEDs using simple stellar populations from \cite{Bruzual03} and a delayed star-formation history with optional late burst and a \cite{Chabrier03} initial mass function. We also  accounted for nebular emission, dust attenuation with a modified \cite{Calzetti00} law, and dust thermal emission using the \cite{Draine14} templates. In particular, CIGALE treats consistently dust attenuation and re-emission, thus conserving the total energy. A summary of the grid values used for SED fitting is reported in Appendix~\ref{appendix_CIGALE}. 

Since we are fitting X-ray selected AGN SEDs, we also included the CIGALE AGN module based on \cite{Stalevski16} and the X-ray module of \cite{Yang19,Yang22}. The X-ray module is especially useful to constrain the AGN component in the SEDs. In fact, bright X-ray emission is largely dominated by the AGN flux, and CIGALE uses the well-known correlation between the AGN intrinsic luminosities at rest-frame UV and X-ray wavelengths \citep[e.g.,][]{Just07} as a prior to constrain the AGN intrinsic optical/UV luminosity. This procedure is particularly valuable in the presence of obscuration, as in this case the AGN optical/UV emission is strongly suppressed. The X-ray module  includes the X-ray emission from binaries and hot gas that depends on the SFR and stellar mass, although it does not include shocks that a starburst can have. These contributions are expected to be negligible at the observed X-ray luminosities \citep[e.g.][]{Lehmer16, Lehmer19}. The fitted X-ray fluxes used in the fitting procedure correspond to the X-ray models discussed in Sec.~\ref{C1_spec_analysis} and \ref{C6_spec_analysis}, and have been corrected for absorption, as required by CIGALE.

The optical photometry of C1 is contaminated by a spectroscopically identified foreground galaxy at $z=2.54$ \citep{Rotermund21}. Following \cite{Hill22}, we thus consider the optical fluxes as upper limits for source C1.  The SED-fitting procedure for this object returns an AGN bolometric luminosity $L_{bol}=(1.9\pm0.7)\times10^{47}\,\lunit$, which is slightly higher than the value estimated in Sec.~\ref{C1_spec_analysis}. The large uncertainties are probably due the fact that only the X-ray point provides an anchor for the AGN component, as only upper limits are used for the optical photometry and the rest-frame mid-IR emission is not sampled by the available datasets.  CIGALE returns a star-formation rate averaged over the last 100 Myr of $\mathrm{SFR_{100Myr}}=228\pm140\,\mathrm{M_\odot\,yr^{-1}}$. This value is significantly lower than the SFR reported by \cite{Hill20}, who estimated it from the FIR luminosity and thus on similar timescales, but did not  account for the AGN component, which can contribute significantly to the total IR luminosity \citep[e.g.][]{DiMascia21a,McKinney21}.  
The stellar mass $\mathrm{M_*}=(3.5\pm2.8)\times10^{10}\,\mathrm{M_\odot\,yr^{-1}}$ is nominally lower than the estimate of \cite{Hill22} , but consistent within the large uncertainties.

Considering C6, we corrected the observed X-ray flux assuming the lower limit we could place on  $N_H$ in Sec.~\ref{C6_spec_analysis}. Thus, the contribution of the AGN component to the best-fitting model for this source might have been underestimated, as higher intrinsic X-ray fluxes would correspond to the observed ones for higher values of $N_H$. We further assumed that the radio emission detected by \cite{Chapman23} in the core region of SPT2349--56 is entirely associated with the X-ray AGN hosted in C6, and thus when fitting this source we considered also the radio fluxes and upper limits from ASKAP, MeerKAT, and ATCA observations presented in that work. This addition required us to employ the CIGALE radio module, that models non-thermal radio emission from star formation and AGN. The CIGALE fit requires a moderately luminous AGN ($L_{bol}=(3.3\pm0.2)\times10^{45}\,\lunit$), which is sub-dominant at all frequencies, except for the X-ray and radio bands. In particular, C6 is a radio-loud\footnote{The radio-loudness parameter is defined in CIGALE as $R = f_{\nu,\mathrm{5GHz}}/  f_{\nu,\mathrm{2500\AA}}$,  i.e. the ratio of the flux densities at rest-frame 5 GHz and 2500 $\mathrm{\AA}$  \citep{Kellerman89, Yang22}.} AGN, as also discussed by \cite{Chapman23}, with $R=98\pm9$.
We obtain $\mathrm{SFR_{100Myr}}=263\pm77\,\mathrm{M_\odot\,yr^{-1}}$
and $\mathrm{M_*}=(3.6\pm1.3)\times10^{10}\,\mathrm{M_\odot\,yr^{-1}}$. Both of these values are significantly lower than those reported by  \cite{Hill20} and  \cite{Hill22}.
In particular, we note that the stellar-mass value that we find is more consistent with the dynamical mass of $\approx(2-5)\times10^{10}\,\mathrm{M_\odot}$ estimated by \cite{Chapman23}.

The stellar masses of these two galaxies correspond roughly to the break mass of the galaxy stellar mass function at $z\approx4$ \citep[e.g.][]{Song16, Weaver23}.
Comparing to the SED fitting results of \cite{Monson23} on X-ray selected AGN in the SSA22 protocluster at $z=3.09$, the AGN in SPT2349--56 on average are hosted in slightly less massive, but significantly more star-forming galaxies. In fact, \cite{Monson23} found that most of the AGN in SSA22 are located below the main sequence, while few of them are consistent with it within the uncertainties on mass and SFR.  Instead, according to our findings, both C1 and C6 are consistent with being main-sequence or even starbursting galaxies  \citep[e.g.][]{Khusanova21, Popesso23}.  The AGN in SPT2349--56 might be in an earlier stage of galaxy evolution than those in SSA22, and are probably still in the peak phase of stellar and BH mass assembly. We also note that  the SFRs of C1 and C6 averaged on a shorter timescale, i.e. 10 Myr, and the instantaneous SFRs returned by CIGALE are even higher than the values reported above (see Tab.~\ref{Tab_SED_results}). 
This is due to the SFHs of these galaxies favoring a recent and short burst of star formation.

\begin{table}
	\caption{Best-fitting physical parameters of C1 and C6 obtain via SED fitting. }

	\begin{tabular}{c|cccccccccc} 
		\hline
			\multicolumn{1}{c|}{{ }} &
			\multicolumn{1}{c}{{ C1 }} &
		\multicolumn{1}{c}{{ C6 }} \\
		\hline
		SFR$_{100\mathrm{Myr}}$ [$\mathrm{M_\odot\,yr^{-1}}$ ]& $228\pm141$& $263\pm77$\\
		 SFR$_{10\mathrm{Myr}}$  [$\mathrm{M_\odot\,yr^{-1}}$ ]&  $1447\pm677$ & $1105\pm198$ \\
		 SFR$_{0\mathrm{Myr}}$  [$\mathrm{M_\odot\,yr^{-1}}$ ]& $2127\pm1292$& $1232\pm242$  \\
		 M$_*$  [ $10^{10}$\,M$_\odot$] &$3.5\pm2.8$ & $3.5\pm1.4$ \\
		 L$_\mathrm{bol,AGN}$  [$\lunit$]&  $(1.9\pm0.7)\times10^{47}$& $(3.3\pm0.2)\times10^{45}$\\
		Radio Loudness & --- & $98\pm9$\\
		\hline
	\end{tabular} \label{Tab_SED_results}\\
	\tablefoot{The rows report the best-fitting SFRs averaged over 100 Myr and 10 Myr, the instantaneous SFRs, the stellar masses, and the AGN bolometric luminosities.}
\end{table}

\subsection{X-ray stacking analysis}\label{stacking}
We performed an X-ray stacking analysis on the protocluster members to check for sub-threshold X-ray emission and constrain the average X-ray luminosity. We used the CIAO \textit{wavdetect} and \textit{dmfilth} tools to identify detected X-ray sources, and replace them with Poisson noise sampled from nearby regions. We then used the \textit{dmcopy} tool to cut thumbnails of the X-ray images and exposure maps  centered at the positions of the protocluster galaxies, excluding C1 and C6 that are detected individually. We summed them separately in the soft, hard, and full bands. We note that all of the protocluster members are within $\approx3$ arcmin from the average aim point of the observations, such that we do not expect a strong variation of the PSF at their positions. The sum of the counts in a $R=2$ pixel (i.e., $\approx1$ arcsec) region around the centers of the stacked images divided by the average values of the stacked exposure maps in the same regions returns the stacked count rates in the three bands. We estimated the background level from nearby regions in the stacked images. We assessed detection significance and net-count numbers following the procedure in \S~\ref{source_detection}.

We did not detect significant stacked X-ray emission from the protocluster members in any energy band. We repeated the procedure considering first all of the possible members (i.e., 97 objects), then only those  identified spectroscopically (i.e., 36 objects), and finally only the DSFGs (i.e., 20 objects). 
Assuming obscured powerlaw emission with $\Gamma=1.9$ and $N_H=10^{24}\,\mathrm{cm^{-2}}$, the lack of significant stacked signal for the sample of DSFGs corresponds to an average intrinsic X-ray luminosity $<2\times10^{43}\,\lunit$. We estimate similar values for the other samples of stacked galaxies. We stress that the stacked exposure times are in the range $5\times10^6$ s to $1.9\times10^7$ s, depending on the stacked sample.
Based on these results, we did not find evidence for widespread low-rate SMBH accretion in the structure, although we cannot exclude that some of the protocluster members host faint AGN.

\begin{table*}
	\caption{Fractions of X-ray selected AGN among different galaxy populations in protoclusters, as described in Appendix~\ref{appendix_fAGN}. }
	\begin{tabular}{c|c|c|cc|cc|cc} 
		\hline
		\multicolumn{1}{c|}{{ Protocluster }} &
		\multicolumn{1}{c|}{{$z$}} &
		\multicolumn{1}{c|}{{ $\fDSFG$}} &		
		\multicolumn{2}{c|}{{ $\fSED$}} &		
		\multicolumn{2}{c|}{{ $\fLAE$}} &		
		\multicolumn{2}{c}{{ $\fHAE$}} \\
		\multicolumn{1}{c|}{} &
		\multicolumn{1}{c|}{} &
		\multicolumn{1}{c|}{spec. } &
		\multicolumn{1}{c}{spec. } &
		\multicolumn{1}{c|}{all} &		
		\multicolumn{1}{c}{spec. } &
		\multicolumn{1}{c|}{all} &		
		\multicolumn{1}{c}{spec. } &
		\multicolumn{1}{c}{all} \\
		\hline
		
		CL 0218.3--0510 & 1.62 & --- & --- & $0.17^{+0.06}_{-0.05}$ & --- & --- & --- & $0.03^{+0.05}_{-0.02}$\\
		Spiderweb & 2.156 & $0.14^{+0.17}_{-0.08}$ & $0.50^{+0.13}_{-0.13}$ & $0.14^{+0.05}_{-0.04}$ & $0.19^{+0.09}_{-0.06}$ & $0.09^{+0.05}_{-0.03}$  & $0.17^{+0.05}_{-0.04}$ &$0.13^{+0.04}_{-0.03}$\\
		PHz G237.01+42.50 & 2.16 & $0.50^{+0.22}_{-0.22}$.  & $0.10^{+0.07}_{-0.04}$ & --- &  ---& --- &$0.17^{+0.20}_{-0.10}$&--- \\
		2QZ Cluster &  2.23 & ---&--- & ---& ---& ---& ---& $0.32^{+0.10}_{-0.09}$\\
		HS 1700+643 & 2.30 & $<0.36$ & $0.05^{+0.05}_{-0.02}$& ---& $0.07^{+0.06}_{-0.03}$ &--- &$0.08^{+0.11}_{-0.05}$& ---\\
		USS 1558--003 &  2.53 &$<0.19$ & ---& ---&--- & ---& $0.04^{+0.03}_{-0.02}$& $0.02^{+0.02}_{-0.01}$\\
		SSA22&  3.09& $0.50^{+0.14}_{-0.14}$& --- & $0.22^{+0.09}_{-0.07}$& --- & $0.03^{+0.02}_{-0.01}$& --- & --- \\
		DRC &  4.002& $0.15^{+0.12}_{-0.07}$& ---& ---&--- &--- & ---& ---\\
		SPT2349--56 &  4.3& $0.09^{+0.04}_{-0.08}$& $<0.36$& ---&$<0.21$ &--- & ---& ---\\
	\end{tabular} \label{Tab_fAGN}\\
	\tablefoot{Uncertainties (upper limits) are at the 68\% (90\%) confidence level, and have been computed as the Jeffrey Bayesian Intervals for binomial proportions \citep[e.g.][]{Brown01}.}
\end{table*}

\section{Discussion}\label{discussion}
In this section we discuss some implications of the results presented in \S~\ref{results}. In particular, in \S~\ref{ISM} we estimate the contribution of the diffuse gas in the AGN host galaxies, i.e., the interstellar medium, to the nuclear obscuration. In \S~\ref{fAGN} we estimate the incidence of AGN among the protocluster member galaxies, and we compare it with other protoclusters and expectations in the field environment. In \S~\ref{AGN_enhancement} we quantify the enhancement of luminous AGN discovered in $z\approx4$ gas-rich protocluster cores.

\subsection{ISM obscuration}\label{ISM}

In the past decades, several works based on X-ray surveys established observationally that the fraction of obscured AGN increases significantly from the local universe up to at least  $z\approx4-5$ \citep[e.g.][]{LaFranca05,Treister06,  Buchner15, Lanzuisi18,  Vito18a,Iwasawa20, Peca23}. 
A possible explanation for that behaviour is that the contribution of the diffuse gas in the host galaxies (i.e., the interstellar medium; ISM) to the nuclear obscuration of AGN increases strongly from the local to the high-redshift universe, due to the larger gas content and smaller sizes characterizing galaxies at early cosmic epochs that almost automatically produce larger gas column densities. This scenario has been tested both observationally and via numerical simulations \citep[e.g.][]{Gilli14,  Circosta19, Trebitsch19, DAmato20, Ni20, Lupi22, Vito22}. In particular, \cite{Gilli22} developed an analytical model for the ISM obscuration that accounts for the ISM clumpiness with a distribution of cloud sizes, masses, and densities, and showed that it predicts well the measured evolution of the obscured AGN fraction up to $z\approx4$.

We tested whether the ISM in the two AGN discovered in SPT2349--56 can contribute significantly to their heavy nuclear obscuration, under simple geometrical assumptions. 
Following Sec. 4.1 of \cite{Gilli22}, we assumed a smooth distribution of the ISM and  that the [C II] surface brightness traces the ISM density, which is distributed in a disk with an exponential profile. This is consistent with the ALMA high-resolution imaging of C1 presented by \cite{Hill22}, in which the galaxy appears as a nearly edge-on disk, and the [C II] light profile is best fitted with a Sersic model with index $n=1.07\pm0.02$, that is, close to an exponential profile. \cite{Gilli22} showed that, in that case,  the gas column density for a line-of-sight inclined by an angle $\theta$ is 
\begin{equation}
	N_H(\theta) = \frac{r_0	\rho_0}{\textit{sin}\theta}(1-e^{-\frac{h}{r_0}\textit{tan}\theta}) 
\end{equation}

where $r_0$ is the scale radius, which for a pure exponential disk can be expressed in terms of the half-light radius $r_{hl}=1.678r_0$, $\rho_0 = \frac{2.8M_\textit{gas}}{4h\pi r_{hl}^2}$ is the central gas density, and $2h$ is the disk thickness. It is further assumed a typical thickness $h=0.15r_{hl}$ \citep[][and references therein]{Gilli22} and no vertical gradient for the gas density. 

For C1, \cite{Hill20} estimated a molecular gas mass of $M_{H_2}=7.5\times10^{10}\,\mathrm{M_\odot}$ from the CO(4--3) emission line, and \cite{Hill22} measured a [C II] half-light radius\footnote{Since the [C II] emission traces the total gas, including the diffuse component, and is thus an upper limit on the extension of the molecular gas. In this sense, our estimate of $N_H$ is conservative. } $r_{hl}=2.91$ kpc. Following \cite{DAmato20}, we consider $M_\textit{gas}=\frac{6}{5}M_{H_2}$ to account for the mass of atomic gas in the galaxy. Therefore,  assuming an edge-on configuration for C1, we obtain $N_H^{\theta=90^\circ}\approx1.2\times10^{24}\,\mathrm{cm^{-2}}$, while assuming the average viewing angle for random inclinations we get $N_H^{\theta=57.3^\circ}\approx4\times10^{23}\,\mathrm{cm^{-2}}$. Smaller inclinations would not be consistent with the nearly edge-on [C II] image of this galaxy, under the assumption of a disk geometry, which is consistent with recent ALMA observations of $4<z<5$ DSFGs \citep[e.g.][]{Rizzo21}. These order-of-magnitude values are close to the column density measured for C1 from our X-ray spectral analysis (Tab.~\ref{Tab_info}), confirming that the gas distributed in the host galaxy can contribute significantly to the nuclear obscuration. We note that the use of the [C II] profile as a proxy of the molecular gas extension is conservative, as that transition traces also atomic gas, which is typically more extended than the dense molecular phase.

The molecular gas mass and [C II] half-light radius of C6 are 
$ M_{H_2}=3.7\times10^{10}\,\mathrm{M_\odot}$ and $r_{hl}=1.30$ kpc \citep{Hill20,Hill22}. Although the [C II] morphology is best fitted with a Sèrsic profile $n=0.78$ \citep{Hill22}, we considered an exponential profile for simplicity. C6 appears close to face-on in the [C II] imaging  \citep{Hill22}, and thus we use $\theta=57.3$ and $\theta=0$ as boundary values of its inclination, finding $N_H^{\theta=57.3^\circ}\approx1.1\times10^{24}\,\mathrm{cm^{-2}}$ and $N_H^{\theta=0^\circ}\approx7\times10^{23}\,\mathrm{cm^{-2}}$. Therefore, also for C6 the nuclear obscuration observed in the X-ray band can be due partly or totally to the gas in the host galaxy. In particular, we note that the ISM in C6 can reach higher column densities than C1 at a given inclination angle because of its compactness, that only high-resolution ALMA imaging could probe.

	We stress that in this section we do not aim to  measure the ISM contribution to the nuclear obscuration estimated via the X-ray observations, but rather to check whether such contribution can in principle be significant.	
	One key assumption of the computations above is that the ISM is distributed smoothly in the galaxies, while it is known to be clumpy. Thus, depending on the geometry and physical properties (e.g., size and mass distribution) of the individual clouds, the ISM column density can be significantly different from, and possibly much lower than, the values estimated in this section. However, due to the large total gas masses estimated for C1 and C6 and
depending on the assumed sizes and masses of molecular gas clouds  \citep[e.g.,][]{Miville-Deschenes17, DessaugesZavadsky19}, the cloud filling factors in these galaxies can be as large as 100\%. Therefore, the assumption of a smooth ISM is reasonable for the order-of-magnitude computations of this section.

\begin{figure*}
		\includegraphics[width=180mm,keepaspectratio]{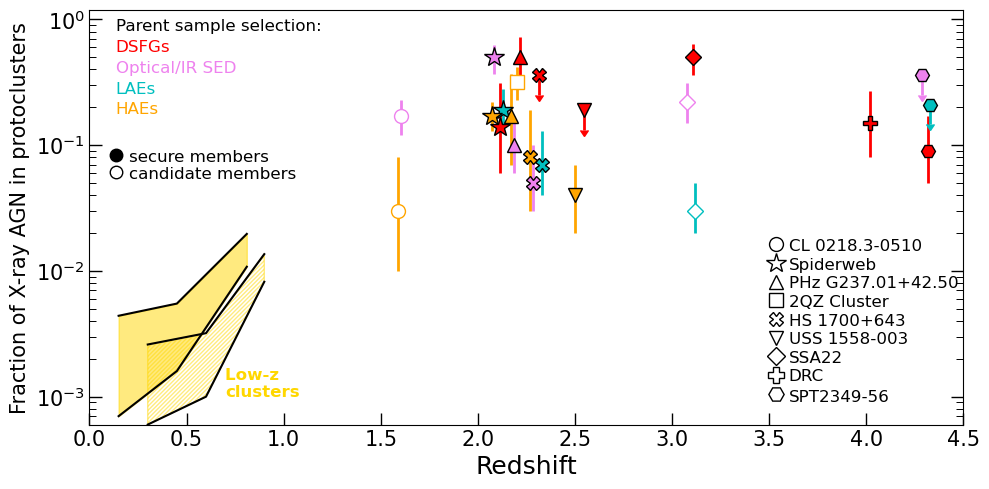} 
	\caption{Fraction of X-ray selected AGN in a sample of protoclusters  as a function of redshift. The different symbol shapes and colors correspond to different structures and selection methods of the parent galaxy population, as marked in the figure. Different symbols for the same structure are slightly shifted in redshift for clarity. 
	We consider the parent samples of spectroscopically confirmed protocluster members when available (filled symbols), otherwise we consider all member candidates (empty symbols).  We refer to Sec.~\ref{fAGN} and Appendix~\ref{appendix_fAGN} for the computation of these values and the relevant citations. For comparison with low-redshift, virialized systems, the filled and hashed yellow stripes represent the X-ray AGN fractions in galaxy clusters presented by \cite{Martini09} and \cite{Bufanda17}, respectively.}\label{Fig_fAGN}
\end{figure*}

\subsection{Incidence of AGN activity in SPT2349--56}\label{fAGN}

We detected two X-ray AGN out of 38 known members,
implying an overall AGN fraction of $0.05_{-0.03}^{+0.05}$, where uncertainties are based on the Jeffrey Bayesian credible interval for binomial proportions \citep[e.g.][]{Brown01}. However, different populations of galaxies are intrinsically characterized by different AGN incidences, such that any comparison with other protoclusters or blank fields should take the specific galaxy selection into account \citep[e.g.][]{Vito23}. Therefore, in the following we consider separately galaxies selected as DSFGs, LAEs, and LBGs.

Two out of the  22  galaxies selected as DSFGs in SPT2349--56 are X-ray selected AGN, corresponding to
an X-ray AGN fraction in such a galaxy population of $0.09_{-0.04}^{+0.08}$. 
This value is remarkably close to the AGN fraction in DRC \citep{Vito20} down to similar FIR luminosity limits, adding the AGN content to other physical properties in common between the two protoclusters, thus suggesting that they have been caught during a similar phase of their galaxy and SMBH evolution.
Since SPT2349--56 and DRC share a similar selection and are located at similar redshifts, we use both of them jointly to improve the number statistics and  estimate the fraction of AGN among DSFGs in $z\approx4$ gas-rich protoclusters by considering all of their DSFG members together (35, four of which are X-ray AGN), finding an X-ray AGN fraction of $0.11_{-0.04}^{+0.06}$.

These results are consistent with the typical ranges found for DSFGs \citep[e.g.][]{Alexander05, Georgantopoulos11, Wang13_1, Shanks21}. However, such samples have typically lower redshift (i.e., $z=2-3$, where the cosmic AGN activity peaks; e.g., \citealt{Aird15}) than SPT2349--56, such that a direct comparison may be misleading. This is due to the quite strong evolution of the cosmic AGN and DSFG populations, which is also dependent on the considered luminosity regimes \citep[e.g.][]{Ueda14, Aird15, Traina24}. For example, if the space density of the X-ray AGN selected AGN decreases from $z=2$ to $z=4$ more strongly than the density of DSFGs, as it appears from observational works, it will cause generally a decreasing AGN fraction at increasing redshift. In this case, observing similar AGN fractions at $z=4$ as at $z=2$ would suggest stronger positive environmental effects at high redshift, but this is currently quite speculative.

In order to factor out the cosmic evolution of the AGN population, the comparison should be made  with a sample of DSFGs at similar redshift as SPT2349--56. We collected a total of 54 sub-mm-selected galaxies in the E-CDFS, COSMOS, and UDS fields from \cite{DaCunha15}, \cite{Scoville16}, and \cite{Dudzeviciute20}, respectively, with photometric redshift $3.5<z<4.5$ and overlapping with the available X-ray coverage in those fields \citep{Civano16,Xue16,Kocevski18}, with sensitivities similar to, or deeper than, the \chandra observations covering SPT2349--56 and DRC.  
We found no match with the X-ray catalogs, corresponding to an observed AGN fraction among field DSFGs $<0.04$. To test quantitatively if the AGN fraction in  $z\approx4$ gas-rich protoclusters is consistent with the field value, we ran the Boschloo's exact test for a 2x2 contingency table. The null hypothesis is that the intrinsic AGN fractions in protoclusters and in the field  are equal, with the observed difference due only to statistical fluctuations. Considering SPT2349--56 only, the test returns a probability of $\approx5\%$ to obtain only by chance a case at least as extreme as the observed ones (i.e., no AGN out 54 DSFGs in the field, and $\geq2$ AGN out of 22 DSFGs in SPT2349--56).  Considering SPT2349--56 and DRC together (i.e., $\geq4$ AGN out of a total of 35 DSFGs), such probability decreases to $\approx1\%$.  This comparison points toward a higher 
AGN incidence among DSFGs in $z\approx4$ protoclusters than in the field at similar redshift,
although the AGN content of additional and similar structures should be investigated to obtain a definitive proof.
Moreover, we caution that the reference samples have been selected differently from observations performed at different sub-mm/mm frequencies and with different depths, and thus might have different flux, luminosity, and mass distributions from the DSFGs population in protoclusters.

Considering the entire sample of 30 protocluster members spectroscopically identified with ALMA via detection of the [C II] and CO(4--3) emission lines, thus including both DSFGs and sources undetected in sub-mm/mm continuum, the AGN fraction decreases slightly to $0.07_{-0.03}^{+0.06}$.
None of the 8 spectroscopically confirmed Ly$\alpha$ emitter galaxies (LAEs) discovered by \cite{Apostolovski23}
 in the SPT2349--56 structure is significantly detected in the X-rays, resulting in an upper limit on AGN fraction among that galaxy population of $<0.21$, consistent with results from blank fields \citep[e.g.][]{ Lehmer09b, Digby-North10, Zheng10}. We obtain no X-ray detection also of the 4 LBGs in the SPT2349--56 core \citep{Rotermund21}, implying an AGN fraction of $<0.36$. 

In Fig.~\ref{Fig_fAGN}, we compare the X-ray AGN fractions of SPT2349--56 with those of a collection of other protoclusters covered with sensitive X-ray observations, as computed in Appendix~\ref{appendix_fAGN} and reported in Tab.~\ref{Tab_fAGN}. The AGN incidence among DSFGs in SPT2349--56 is consistent with the results for other $z>2$ protoclusters, although a couple of structures have significantly higher AGN fractions. Instead, the AGN fraction  drops dramatically in virialized clusters at lower redshift. This behavior might be connected to the virialization of the structures or strong AGN feedback hindering the infall of large amounts of cold gas into galaxies, and thus the triggering of luminous nuclear activity.  We note that the points in Fig.~\ref{Fig_fAGN} are observed fractions, and we refer to Appendix~\ref{appendix_fAGN} for a discussion of the caveats. A more in-depth investigation of the possible cosmic evolution of the AGN incidence in protoclusters would require taking several effects into account, among which are the different sensitivities of the multi-wavelength observations, the dependence of SMBH accretion on the host-galaxy stellar mass, and the intrinsic cosmic evolution of the AGN population \citep[e.g.,][]{Aird15, Aird18, Yang17, Yang18,Zou24}, which can be controlled for by comparing with the field AGN incidence. 
{Such analysis require, among other things, a proper assessment of the multi-band observation sensitivities across multiple extragalactic fields and of the different specific selections applied on such fields, as well as a consistent SED fitting analysis of the resulting large samples of SMGs. These tasks are beyond the scope of this paper, and we reserve it for a dedicated future work.}

\subsection{Enhancement of fast SMBH growth in $z\approx4$ overdensities of DSFGs}\label{AGN_enhancement}

The obscuration level and luminosity of C1 are remarkably similar to those of DRC-2 (Fig.~\ref{Fig_C1_Lx_NH}).  To our knowledge,
AGN with similar X-ray luminosities in protocluster environments have been detected only in two structures at $z=2.16-2.41$ by \cite[see Fig.~\ref{Fig_C1_Lx_NH}]{Ivison19, Tozzi22b}, and those are unobscured  or at most mildly obscured objects.
The detection of a luminous, Compton-thick AGN  in the core regions of the only two $z\approx4$ protoclusters selected as overdensities of dusty star-forming galaxies and covered by sensitive X-ray observations suggests that gas-rich and dense regions of the Universe at those epochs may promote the triggering of extremely fast SMBH growth in heavily obscured conditions.

\begin{figure}
		\includegraphics[width=90mm,keepaspectratio]{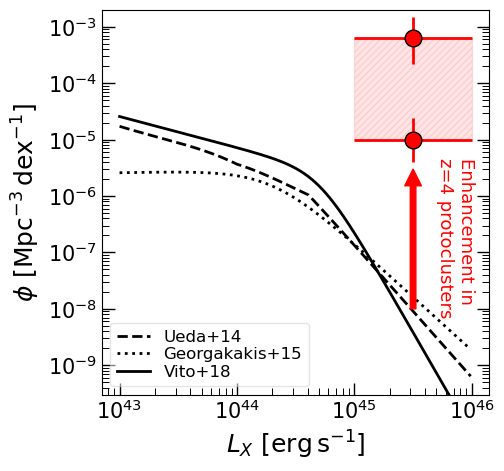} 
	\caption{Space density of luminous (log$\frac{L_X}{\lunit}=45-46$) and obscured AGN in gas-rich protocluster cores at $z=4.0-4.3$ (red circles) computed under two assumptions for the volumes of the structures (see Sec.~\ref{AGN_enhancement}), compared with the predictions of AGN X-ray luminosity functions at $z=4.15$ in blank fields \citep{Ueda14,Georgakakis15,Vito18a}. Gas-rich overdense environments at high redshift enhance the triggering of luminous AGN by 3--5 orders of magnitude. This is likely a physical effect, as it appears not to be simply driven  by the large number of galaxies in the structures (see \S~\ref{AGN_enhancement}). }\label{Fig_space_density}
\end{figure}

To estimate the level of enhancement of luminous AGN in SPT2349--56 and DRC, we compare their space density with that of AGN in the field environment with similar luminosity and redshift.  We assume that the two protoclusters are enclosed in spherical volumes with radii equal to the projected distances between the observed centers of the structures and the farthest spectroscopically confirmed members, which are $~\approx8.8$ comoving Mpc ($\,\mathrm{cMpc}$) and $~\approx3.8 \,\mathrm{cMpc}$, respectively \citep{Ivison20,Hill22}. Accounting for the uncertainties on their estimated luminosities, we consider the two luminous AGN in SPT2349--56 and DRC as representative of the AGN population in the range log$\frac{L_X}{\lunit}=45-46$ in $z\approx4$ gas-rich protoclusters. The space density of such a population is then two divided by the sum of the volumes computed above; i.e., $\Phi_{\mathrm{AGN}}^{\mathrm{prot}}=6.4_{-4.2}^{+8.5}\times10^{-4}\,\mathrm{cMpc^{-3}\,dex^{-1}}$, where the uncertainties account for the statistical errors on the number of objects \citep[]{Gehrels86}.
The space density of log$\frac{L_X}{\lunit}=45-46$ at $z=4.15$ in the field environment is
 $\Phi_{\mathrm{AGN}}^{\mathrm{field}}\approx5\times10^{-8}\,\mathrm{cMpc^{-3}\,dex^{-1}}$ \citep[Fig.~\ref{Fig_space_density}; e.g.,][]{Gilli07, Ueda14, Aird15, Vito18a}, corresponding to an expected number of $1.5\times10^{-4}$ luminous AGN in the considered volume.
 The Poisson probability of instead finding two AGN with such luminosity by chance only is negligible.  This simple computation suggests that the triggering of luminous AGN in gas-rich protocluster environments at $z\approx4$ is enhanced by about four orders of magnitude with respect to the field environment at similar redshift with high significance. 

The most uncertain quantities that enter in the estimate are the volumes of the two protoclusters. As a second and more conservative estimate, we assumed that these structures extend up to $R=28\,\mathrm{cMpc}$. According to \cite{Muldrew15}, this is the average radius that encloses 90\% of the stellar mass of protoclusters at $z=4$ that form massive galaxy clusters at $z=0$. In this case, we estimate $\Phi_{\mathrm{AGN}}^{\mathrm{prot}}=1.0_{-0.6}^{+1.4}\times10^{-5}\,\mathrm{cMpc^{-3}\,dex^{-1}}$, which is still $>2$ dex higher than the field value.  The Poisson probability of finding two luminous AGN while expecting the number predicted by the field environment is negligible also in this case.

As a comparison, \cite{Tozzi22a} found for the Spiderweb protocluster at $z\approx2$ an AGN enhancement of a factor of tens, depending on the AGN luminosity. This lower value might be due to the fact that $z\approx2$ protoclusters have often already consumed most of their gas, as suggested by the large fraction of passively evolving galaxies in the Spiderweb structure \citep{Shimakawa24}, and to the overall cosmic evolution of the AGN space density, that peaks close to that epoch.

The much higher space density of luminous AGN in SPT2349--56 and DRC than in the field at similar redshift can in principle be driven by the space density of the underlying galaxy population, which in protoclusters is enhanced with respect to the field by definition. For instance, \cite{Miller18} estimated a SMG overdensity in SPT2349--56 of a factor $>1000$. However, they considered only the central $R=130$ kpc region, where their selection is complete at $S_{1.1\,\mathrm{mm}}>0.5\,\mathrm{\mu Jy}$. In that same region we detected C1, i.e., $1.0_{-0.8}^{+2.3}$ X-ray luminous AGN, which is at least several million times larger than the expected number of $4\times10^{-8}$ similar objects in the field, assuming a spherical volume with that radius.  This enhancement largely exceeds and thus is hardly driven by the overdensity level of the underlying galaxy population in the considered protoclusters. We stress that  C1 is not an extreme galaxy in terms of stellar mass  (see \S~\ref{SED_fitting}), and thus the comparison with the  overdensity estimated by \cite{Miller18} for the entire population of DSFGs in the structure is fair.  This result is only in apparent contrast with the lower significance found for the enhanced AGN fraction in protoclusters with respect to the field environment discussed in \S~\ref{fAGN}, since in this section we focused on the high-luminosity regime only, and compared space densities rather than AGN fractions. Given the shape of the X-ray luminosity function (Fig.~\ref{Fig_space_density}), such luminous AGN are extremely rare in the field, and finding two of them in small volumes corresponds to a large enhancement factor. In fact, the small enhancement of the overall AGN fraction coupled with the large enhancement of the space density of high-luminosity AGN suggests that the AGN X-ray luminosity function in these protoclusters is flatter than in the field, as also found by \cite{Tozzi22a} for the Spiderweb protocluster. 
These computations suggest  that gas-rich overdensities of DSFGs  at $z\approx4$ promote extremely luminous and obscured AGN activity. A larger sample of similar protoclusters is needed to confirm it with better statistics.

We note that \cite{Yang18b} investigated the possible dependence of the average SMBH accretion rate density in samples of galaxies in the COSMOS field as a function of their environments up to 10 Mpc, finding no significant difference in the SMBH accretion power between overdense and field regions at a fixed stellar mass. However, they analysis is limited to $z<3$ and  the environments that they probed are not as dense as SPT2349--56.

The luminous AGN in the $z\approx4$ protocluster cores represent the phase of fast SMBH growth required to explain the masses of SMBHs in the centers of low-redshift galaxy clusters. 
Such AGN are likely caught just before the "blow-out" phase, when AGN feedback clears the line of sight of most of the obscuring material \citep[e.g.][]{Ivison19}, and eventually hinders star formation and further SMBH growth.  X-ray observations of a larger sample of similar environments are required to investigate the AGN population in such structures and prove this scenario securely. The specific selection that led to the identification of SPT2349--56 and DRC appears to be key to identifying high-redshift structures hosting such luminous AGN. In fact, other protoclusters selected as overdensities of Lyman-break galaxies via optical observations do not present strong evidence for the presence of such an AGN population, although they host an unusual large number of rest-frame UV bright galaxies \citep{Toshikawa24}.

\section{Summary and conclusions}\label{conclusions}

We presented new \chandra observations of the $z=4.3$ SPT2349--56 protocluster, which was identified as an extreme overdensity of DSFGs \citep{Miller18,Hill20,Rotermund21,Hill22,Chapman23}. We summarize here our main results.

\begin{itemize}

\item We identified two X-ray detected AGN among the SPT2349--56 member galaxies, namely C1 and C6, which are among the most gas-rich galaxies in the system \citep{Hill20}. We did not detect significant emission by stacking the X-ray data of the individually undetected galaxies, implying an average X-ray luminosity $<2\times10^{43}\,\lunit$. See \S~\ref{data_analysis} and \S~\ref{stacking}.

\item C1 is an extremely luminous ($L_{2-10\,\mathrm{keV}}=2\times{10^{45}}\,\lunit$), Compton-thick ($N_H=2\times10^{24}\,\mathrm{cm^{-2}}$) AGN. The X-ray luminosity translates into a bolometric power $\approx10^{47}\,\lunit$, which is also confirmed via SED fitting. Assuming that the SMBH accretion is capped at the  Eddington limit, we place a lower limit on its mass of $\approx7\times10^8\,\mathrm{M_\odot}$. Both its luminosity and obscuration level are similar to those of another AGN  previously detected in the central region of DRC, a similar protocluster at $z=4$ \citep{Vito20}. Both of these AGN might have already accreted a significant fraction of the typical mass of SMBHs in the centers of local clusters at much later ($> 10$ Gyr) cosmic times. See \S~\ref{C1_spec_analysis}.

\item Due to the low number of detected X-ray photons, we can only place lower limits on the luminosity ($L_{2-10\,\mathrm{keV}}>2\times{10^{43}}\,\lunit$) and column density ($N_H=4\times10^{23}\,\mathrm{cm^{-2}}$) of C6, which is also a radio-loud AGN. See \S~\ref{C6_spec_analysis}.

\item Both C1 and C6 are hosted in galaxies with stellar masses $\approx3\times10^{10}\,\mathrm{M_\odot}$, which is close the break value of the galaxy stellar mass function at $z=4.3$, and have star formation rates consistent with, or in excess of, the expectation of the main sequence of star-forming galaxies at that redshift. See \S~\ref{SED_fitting}.

\item Under simple, but reasonable, assumptions on the geometries of the host galaxies, we conclude that the ISM can in principle contribute significantly to the observed nuclear obscuration of both AGN, in agreement with previous works on high-redshift AGN, although that contribution can be lower in the case of highly clumpy medium. See \S~\ref{ISM}.

\item The X-ray AGN fraction among DSFGs in SPT2349--56 is about 10\%, consistent with other $z>2$ protoclusters, and in particular with DRC. The fraction is higher than the X-ray AGN incidence in DSFGs in the field environment at $z\approx4$. We could place only loose upper limits on the AGN incidence in LBGs and LAEs in SPT2349--56, due to their small number. See \S~\ref{fAGN}.

\item Both SPT2349--56 and DRC, which share similar selection and physical properties, host highly luminous Compton-thick AGN, indicating the existence of a tight link between vigorous phases of star formation, fed by the availability of huge gas reservoirs, and high SMBH accretion rates in the densest environments at high redshift. Such luminous AGN probably represent the period of fast SMBH growth required to explain the presence of $10^9-10^{10}\,\mathrm{M_\odot}$ SMBHs in the central galaxies of local clusters.
Under different assumptions about the volumes of these structures and comparing with the predictions of the X-ray luminosity function $z=4$, we suggest that gas-rich and dense protoclusters at $z\approx4$ enhance the triggering of extremely fast SMBH accretion by a factor of 3--5 dex with respect to the field environment. This factor exceeds the galaxy overdensity level of the protoclusters,  and thus is probably not merely driven by the large number of galaxies in the structures. 
 Further X-ray observations of similar structures are needed to confirm this result. See \S~\ref{AGN_enhancement}.
 
\end{itemize}

Our results demonstrate that sensitive X-ray observations with high angular resolution are crucial to identify AGN in high-redshift protoclusters, which are characterized by large amounts of dust and gas, and thus heavy nuclear obscuration. In the next years, \chandra 
will play a leading role in this respect, by increasing the samples of high-redshift gas-rich protoclusters with the deep X-ray coverage required to investigate their AGN content. Future X-ray missions will then be crucial to obtain a complete view of the relation between overdense environments and SMBH growth at high redshift \citep[e.g.,][]{Vito23}.
 
 \begin{acknowledgements}
 	We thank the anonymous referee for their useful comments and suggestions.
 	FV thanks R. Hill and S. Chapman for kindly providing the ALMA and ATCA data and for useful discussion. FV acknowledges support from the "INAF Ricerca Fondamentale 2023 -- Large GO" grant.  WNB acknowledges support from CXC grant GO2--23074X.
 	  This research has made use of data obtained from the Chandra Data Archive (Proposal ID 23700087), and software provided by the Chandra X-ray Center (CXC) in the application packages CIAO. This research made use of Astropy, a community-developed core Python package for Astronomy \citep{Astropy13,Astropy18}, and the Statsmodels package \citep{statsmodels}.
\end{acknowledgements}

\appendix

\section{Grid of parameters value used for the SED fitting}\label{appendix_CIGALE}

Tab.~\ref{Tab_CIGALE} lists the parameters and values used for the SED fitting procedure with CIGALE, as described in Sec.~\ref{SED_fitting}. 

\begin{table*}
	\caption{Parameters and values for the modules used with CIGALE. Parameters not listed here are fixed to their default values.}
	\centering
	\setlength{\tabcolsep}{1.mm}
	\begin{tabular}{cc}
		\hline
		Parameter &  Model/values\\
		\hline\\[-1.5ex]
		\multicolumn{2}{c}{Star formation history: delayed model and recent burst} \\[0.5ex]
		Age of the main population & 250, 500 Myr \\
		e-folding time & 100, 250, 500 Myr \\
		Age of the burst & 5, 10, 25, 50 Myr \\
		e-folding time of the burst$^a$ & 10000 Myr \\
		Burst stellar mass fraction & 0, 0.1, 0.3, 0.5, 0.7, 0.9, 1.0 \\
		\hline\\[-1.5ex]
		\multicolumn{2}{c}{Simple Stellar population: \cite{Bruzual03}} \\[0.5ex]
		Initial Mass Function & \cite{Chabrier03}\\
		Metallicity & 0.008, 0.02 (Solar) \\
		\hline\\[-1.5ex]
		\multicolumn{2}{c}{Nebular emission} \\[0.5ex]		
		Gas metallicity & 0.008, 0.014\\
		f$_\mathrm{dust}$ & 0, 0.25, 0.5, 0.75, 1.0\\
				\hline\\[-1.5ex]
		\multicolumn{2}{c}{Galactic dust extinction} \\[0.5ex]
		Dust attenuation  & modified \cite{Calzetti00}\\
		E(B-V)$_\mathrm{lines}$ &   0.1, 0.3, 0.5, 1, 1.5, 2.0\\
		Scale factor to E(B-V)$_\mathrm{stars}$ & 1 \\
		Power-law slope & $-1$, $-0.75$, $-0.5$, $-0.25$, 0\\
		Extinction law & SMC\\
		\hline\\[-1.5ex]
		\multicolumn{2}{c}{Galactic dust emission: \cite{Draine14}} \\[0.5ex]
		$u_\mathrm{min}$ & 2.0, 5, 10, 30, 50 \\
		gamma & 0.02, 0.1, 0.25, 0.5, 0.75 \\		
		\hline\\[-1.5ex]
		\multicolumn{2}{c}{AGN module: SKIRTOR} \\[0.5ex]
		Angle between the equatorial plan and edge of the torus & $60^{\circ}$ \\
		Viewing angle  &$90^{\circ}$ \\
		AGN fraction & 0.0, 0.1, 0.25, 0.5, 0.75, 0.9, 0.99 \\
		$E(B-V)$ of polar dust & 0.1\\
		\hline\\[-1.5ex]
		\multicolumn{2}{c}{X-ray module} \\[0.5ex]
		AGN photon index $\Gamma$ & 1.9 \\
		$\alpha_{ox}$ & -2.0, -1.9, -1.8, -1.7, -1.6, -1.5\\
		\hline\\[-1.5ex]
		\multicolumn{2}{c}{Radio module:$^b$} \\[0.5ex]
	R$_\mathrm{AGN}$ & 10, 25, 50, 75, 100, 125\\
$\alpha_\mathrm{AGN}$ & 1.45 \\

		\hline
		\label{Tab_CIGALE}
	\end{tabular}
	\tablefoot{$^a$ Using an e-folding time of the star-formation burst much higher than its age effectively reproduces a constant burst of star formation over a period equal to the burst age. $^b$ The Radio module is used only for C6 (see Sec.~\ref{SED_fitting}).
	}
\end{table*}

\section{AGN fraction in protoclusters}\label{appendix_fAGN}

We report here the computations used to estimate the X-ray selected AGN fractions $f_\mathrm{AGN}$ among members of a few protoclusters for which dedicated X-ray observations have been obtained and published (Fig.~\ref{Fig_fAGN}). We consider different galaxy populations, based on their selections: sub-mm selected dusty star-forming galaxies (DSFGs), objects selected on the basis of their optical/IR SED (e.g., Lymam-break or BX/MD galaxies), Ly$\alpha$ emitters (LAEs), and H$\alpha$ emitters (HAEs).  We also consider separately the secure members of a protocluster (i.e., those identified spectroscopically) and all possible members, including candidate members, except for the DSFG population, for which only spectroscopically identified galaxies are considered. The resulting AGN fractions are summarized in Tab.~\ref{Tab_fAGN}.

The intention of this collection is to provide an easy-to-access list of values of AGN fractions in the literature, for first-order comparisons. Several caveats should be considered when using these values, some of them are discussed here. First, we consider these structures as \textit{bona fide} protoclusters, based on the definitions provided by the referenced papers. Second, the selection methods (e.g., definitions of colors, thresholds, sampled rest-frame wavelengths, etc.) and sensitivities of the multi-wavelength observations, including the X-ray coverage used to identify AGN,  are not homogeneous among these objects. Third, the fractions have been computed simply considering the number of X-ray detected AGN divided by the number of known galaxies belonging to a given class. A more reliable and complete procedure would take into account, for instance, the varying sensitivities of the X-ray and multi-wavelength observations on a field, and control for the stellar mass distribution of the parent populations of galaxies. A few of the works mentioned below did take these precautions into account for individual protoclusters, but extending such procedures homogeneously to the entire protocluster sample considered here is beyond the scope of this paper. 
Finally, we rely on the published X-ray-to-multiwavelength counterpart matching,  unless otherwise noted. The matching procedures generally vary among different works. As a consequence, the following list is far from being complete and homogeneous.

\textit{CL 0218.3--0510} ($z=1.62$): \cite{Krishnan17} considered 46 massive (i.e., $M_*>10^{10}\,\mathrm{M_\odot}$) galaxies identified as robust protocluster member candidates via optical/IR SED fitting by \cite{Hatch16}, and found that 8 of them are detected in the X-ray band, corresponding to an AGN fraction of 
 $\fSED = 0.17^{+0.06}_{-0.05}$. SMGs have been identified as potential protocluster members \citep[e.g.,][]{Smail14, Chen16}, but spectroscopic confirmation is not available to our knwoledge, so we do not consider this population. \cite{Tran15} reported that 1 out of 33 $H_\alpha$ emitters associated with the protocluster is X-ray detected, corresponding to $\fHAE = 0.03^{+0.05}_{-0.02}$.

 \textit{Spiderweb protocluster} ($z=2.156$). \cite{Tozzi22a} studied the X-ray selected AGN in this structure using deep (700 ks) \textit{Chandra} observations. They reported the detection of 14 protocluster members (13 of which identified spectroscopically), corresponding to an AGN fraction of $f_{AGN}=0.25\pm0.04$ (including systematic uncertainties) among spectroscopically identified, massive (log$\frac{M_*}{M_\odot}>10.5$) protocluster members. Here, we instead report the X-ray AGN fractions corresponding to the different galaxy selection methods, considering no threshold in stellar mass. The galaxy parent populations are collected from the same catalogs used by \citep[][see their Tab. 2]{Tozzi22a}, to which we added the spectroscopic sample recently presented by \cite{PerezMartinez23a}. We obtained AGN fractions of $\fDSFG=0.14^{+0.17}_{-0.08}$, $\fLAE=0.09^{+0.05}_{-0.03}$ , $\fHAE=0.13^{+0.04}_{-0.03}$, and  $\fSED=0.14^{+0.05}_{-0.04}$. All DSFGs are spectroscopically identified, while considering only spectroscopically confirmed objects for the other populations we obtained 
 $\fLAE=0.19^{+0.09}_{-0.06}$,  $\fHAE=0.17^{+0.05}_{-0.04}$, and $\fSED=0.50^{+0.13}_{-0.13}$. We note that Jin+21 selected 46 protocluster members  as CO-emitters. Among this population we found $f\mathrm{^{CO}_{AGN}}=0.09^{+0.05}_{-0.03}$.

 \textit{ PHz G237.01+42.50} ($z=2.16$). \cite{Polletta21} reported the identification of this protocluster with 31 spectroscopic members detected in optical/IR observations. Among them, three are known X-ray AGN, corresponding to $\fSED=0.10^{+0.07}_{-0.04}$. We note that \cite{Polletta21} included also an X-ray undetected broad emission-line AGN in their computation of the AGN fraction in the structure, while here we limited to X-ray selected AGN. Among the 31 members of the protocluster, 6 were also selected as HAEs, one of which is an X-ray AGN, i.e., $\fHAE=0.17^{+0.20}_{-0.10}$. Finally, 2 out of the  4 DSFGs selected with Herschel/SPIRE  among the members are X-ray AGN, corresponding to $\fDSFG=0.50^{+0.22}_{-0.22}$.
 
 \textit{2QZ Cluster} ($z=2.23$). \cite{Lehmer13} presented the \chandra observations of this structure. Seven out of 22 HAEs are X-ray detected (i.e., $\fHAE=0.32^{+0.10}_{-0.09}$), including the four optically selected AGN that were used as signposts of the protocluster. 
 
 \textit{QSO HS 1700+643} ($z=2.30$). Among the spectroscopically confirmed members of the protocluster which do not reside in regions affected by high X-ray background due to the presence of foreground clusters, \cite{Digby-North10} reported the X-ray detection of 2 out of 29 LAEs ($\fLAE=0.07^{+0.06}_{-0.03}$), 1 out of 12 HAEs  ($\fHAE=0.08^{+0.11}_{-0.05}$), and 2 out of 39 BX/MD selected galaxies ( $\fSED=0.05^{+0.05}_{-0.02}$). We note that these fraction are different from those reported by \cite{Digby-North10}, because they considered also objects not confirmed spectroscopically and applied corrections for the X-ray sensitivity over the field. \cite{Lacaille19} identified spectroscopically four DSFGs as members of the protocluster, and none of them are detexcted in the X-rays, corresponding to $\fDSFG<0.36$.

 \textit{USS 1558–003} ($z=2.53$). Combining the catalogs of \cite{Shimakawa18}, \cite{Aoyama22}, and \cite{PerezMartinez23b},  57 HAEs are secure members of this structure, either because of spectroscopic identification or matching detection in narrow-band imaging targeting the Ly$\alpha$ emission line (but we note that the catalog of LAEs has not been published to our knowledge). Two of them are detected in the X-rays \citep{Macuga19}; i.e., $\fHAE=0.04^{+0.03}_{-0.02}$. Considering also member candidates, the fraction is $\fHAE=0.02^{+0.02}_{-0.01}$. 
 Nine DSFGs have been spectroscopically identified in the structure \cite{Aoyama22}, and all of them are also selected as HAEs. Matching their positions with the catalog of \cite{Macuga19}, we found no X-ray detection; i.e.,  $\fDSFG<0.19$.
 
  \textit{SSA22} ($z=3.09$). We base the computation of the X-ray AGN fraction in this protocluster on the work of \cite[see also \citealt{Monson23}]{Lehmer09b}, who reported X-ray detection of six out of 27 member candidates selected as LBGs, i.e., $\fSED=0.22^{+0.09}_{-0.07}$, and five out of 144 LAEs candidates, i.e., $\fLAE=0.03^{+0.02}_{-0.01}$. These fractions are slightly different than those reported in \cite{Lehmer09b}, due to the different way in which they are computed. 
  \cite{Umehata19} presented deep ALMA observations of the SSA22 fields. Among the 12 DSFGs spectroscopically confirmed as members of the protocluster, 6 are detected in the X-ray catalog of \cite{Lehmer09a}; i.e., $\fDSFG=0.50^{+0.14}_{-0.14}$.  We also note that since the publication of those works a few new relevant datasets have been presented \citep[e.g.][]{Yamada12, Kubo15, Kubo16, Topping16, Radzom22}, but are not considered here.
 
 \textit{DRC} (z=4.002). Among the 13 spectroscopically identified SMGs in this structure \citep{Oteo18, Ivison20}, two are X-ray selected AGN, corresponding to $\fDSFG=0.15^{+0.12}_{-0.07}$ \citep{Vito20}.

%
%
\bibliographystyle{aa}
\bibliography{biblio.bib} 
%
%

\end{document}